\documentclass[12pt]{article}

\usepackage[body={17.5cm, 22.5cm},right=2cm]{geometry}

\usepackage{color}
\usepackage{graphicx}
\usepackage{hyperref}
\usepackage{epsf}
\usepackage{graphicx,epsfig}
\usepackage[skip=2pt,font=small]{caption}

\usepackage{float}

\usepackage{amsmath}
\usepackage{amssymb}
\usepackage{epsfig}
\usepackage{cite}
\usepackage{color,colordvi}
\newcommand{\be}{\begin{eqnarray}}
\newcommand{\ee}{\end{eqnarray}}
\newcommand{\bi}{\begin{itemize}}
\newcommand{\ei}{\end{itemize}}

\newcounter{hran}

\def\MSbar{\relax\ifmmode\overline{\rm MS}\else{$\overline{\rm MS}${ }}\fi}

\def\d{\rm d}

\def\d{{\rm d}}

\def\n3{\nu_3}

\def\n1{\nu_1}

\begin{document}

\begin{flushright}
LMU-ASC 06/15, MPP-2015-21, CERN-PH-TH-2015-026, LPTENS-15/01, 
February 2015
\vspace{-.0cm}
\end{flushright}

\vspace*{1cm}
\begin{center}

\def\thefootnote{\fnsymbol{footnote}}

{\Large \bf 
Black Hole Solutions in $R^2$ Gravity
}
\\[1.5cm]
{\large  Alex Kehagias$^{a,b}$, Costas Kounnas$^{c}$, Dieter L\"ust$^{d,e,f}$  and Antonio Riotto$^{b}$}
\\[0.5cm]

\vspace{.3cm}
{\normalsize {\it  $^{a}$ Physics Division, National Technical University of Athens, \\15780 Zografou Campus, Athens, Greece}}\\

\vspace{.3cm}
{\normalsize { \it $^{b}$ Department of Theoretical Physics and Center for Astroparticle Physics (CAP)\\ 24 quai E. Ansermet, CH-1211 Geneva 4, Switzerland}}\\

 \vspace{.3cm}
{\normalsize { \it $^{c}$
Laboratoire de Physique Th\'eorique, Ecole Normale Sup\'erieure,\\
24 rue Lhomond, 75231 Paris cedex 05, France}}\\

\vspace{.3cm}
{\normalsize { \it $^{d}$
Max-Planck-Institut f\"ur Physik (Werner-Heisenberg-Institut),\\
F\"ohringer Ring 6, 80805 M\"unchen, Germany}}\\

\vspace{.3cm}
{\normalsize { \it $^{e}$
Arnold Sommerfeld Center for Theoretical Physics,\\
LMU, 
Theresienstr. 37, 80333 M\"unchen, Germany}}\\

\vspace{.3cm}
{\normalsize { \it $^{f}$
CERN, PH-TH Division\\
1211 Geneva 23,
Switzerland}}

\vspace{.3cm}


\end{center}

\vspace{1cm}

\hrule \vspace{0.3cm}
{\small  \noindent \textbf{Abstract} \\[0.3cm]
\noindent 
We find static spherically symmetric solutions of scale invariant 
${R}^2$  gravity. The latter has been shown to be  equivalent to General Relativity with a positive cosmological constant and a scalar mode. Therefore, one expects that solutions of the $R^2$ theory will be identical to that of Einstein theory. Indeed, we find that the solutions of $R^2$ gravity  are in one-to-one correspondence with solutions of General
Relativity in the case of non-vanishing Ricci scalar. 
However, scalar-flat $R=0$ solutions are global minima of the $R^2$ action and they cannot in general be mapped to solutions of the Einstein theory. As we will discuss,
the $R=0$ solutions arise in Einstein gravity as solutions in the tensionless, strong coupling limit $M_P\rightarrow 0$.
As a further result, there is no corresponding Birkhoff theorem  and the  Schwarzschild black hole  is by no means unique in this framework. In fact, $R^2$ gravity  has a rich structure of  vacuum static spherically symmetric solutions partially  uncovered here. 
We also find charged static spherically symmetric backgrounds coupled to a $U(1)$  field.  
 Finally, we provide  the entropy and energy formulas for the $R^2$ theory and we find that entropy and energy  vanish for scalar-flat backgrounds.

\vspace{0.5cm}  \hrule
\vskip 1cm

\def\thefootnote{\arabic{footnote}}
\setcounter{footnote}{0}


\baselineskip= 15pt

\newpage 

\section{Introduction}

Gravity theories based on Einstein gravity plus higher derivative terms of the type $R+{\cal  R}^n$ possess several interesting features attracting  the attention of our scientific community;  $R$ denotes the scalar curvature and ${\cal R}$ stands for all possible combinations and versions of the curvature tensor.  A first remark is that 
 in string theory an infinite series of higher curvature corrections to Einstein gravity appears naturally. Secondly, as it was noticed in Ref. \cite{Stelle},
the addition of  the quadratic curvature terms with $n=2$ makes the theory of gravity renormalizable. However the prize one has to pay is the appearance of a ghost-like modes in the theory, which are originated from the square of the Riemann tensor as well as from the square of the Weyl tensor. This pathology however is absent once we are restricted to the case of  $R+{ R}^2$; this theory turns out to be ghost-free and it is equivalent to standard Einstein gravity  with one additional scalar degree of freedom $\phi$. Furthermore as was noticed by Starobinsky and others 
\cite{Starobinsky} the potential of $\phi$ is  of great cosmological interest, since it can describe a slow transition from a de Sitter phase to a flat Minkowski phase leading  to a viable 
 realization of the inflationary scenario in the early Universe \cite{Guth}.  More recently the $R+{ R}^2$ inflationary scenario was also investigated  in the context of supergravity 
 \cite{Ellis:2013xoa,Kounnas:2014gda}.

 \vskip0.2cm
In almost all previous investigations of higher curvature gravity and its solutions \cite{Jacobson:1993xs} the linear Einstein term was assumed to be always present in the action. In this work we investigate in more detail the 
case of pure  ${ R}^2$ theory. As it was recently emphasized \cite{Kounnas:2014gda} and also further investigated in the context of supergravity and superstring theory, the pure   ${ R}^2$
gravity possesses two distinct and important features: first it is the only ghost free theory of quadratic order in the curvature tensor. Second, one obtains in this way a scale invariant gravitational theory with a non trivial interacting structure, in analogy to the scale invariance of the low energy (Supersymmetric)-Standard Model without Higgs mass term at the classical level.\footnote{The scale violating terms in particular the Einstein term $R$ will be induced at the quantum level of the theory, where conformal and scale invariance are broken in a more or less controllable way \cite{Kounnas:2014gda}.}
Reformulated in Einstein frame, the ${ R}^2$ theory
describes an action with cosmological constant, i.e. 
an exact de Sitter  (or Anti de Sitter) space as a consequence of an unbroken global scale symmetry. 

\vskip0.2cm

The structure of this paper is as follows: In the next section 2 we construct vacuum static spherically symmetric solutions of the pure $R^2$ gravity. In section 3 we couple the theory to Maxwell theory and we find the corresponding solutions. In section 4 we discuss entropy and energy issues of the background found. Finally we conclude in section 5.

\section{Spherically Symmetric  Static  Backgrounds in $R^2$ gravity}

The most general quadratic and scale invariant theory for gravity is described by the Weyl-Eddington action,
\begin{eqnarray}
S_W=\int d^4 x\sqrt{-g}\Big{(}\alpha C_{\mu\nu\rho\sigma}C^{\mu\nu\rho\sigma}+\beta R^2\Big{)} \label{W}\, ,
\end{eqnarray}
where $C_{\mu\nu\rho\sigma}$ is the Weyl tensor and $R$ is the Ricci scalar (see \cite{Salvio:2014soa}   for some more recent discussions on this action).

For $\beta=0$, the theory describes Weyl gravity which has extensively been studied. In this particular case,  the action (\ref{W}) is not only scale invariant but exhibits full conformal invariance; it is however a problematic theory since it contains ghost degrees of freedom. On the other hand the  limit  $\alpha=0, \beta\ne0$ is ghost free theory and it deserves special  attention; the  pure $R^2$ gravity is described by a simple quadratic action \cite{Kounnas:2014gda},
\begin{eqnarray}
S=\int d^4 \sqrt{-g}\left(\frac{1}{16 \mu^2}R^2+{\cal{L}}_{\rm matter}\right)\, ,
\label{S}
\end{eqnarray}
where  ${\cal{L}}_{\rm matter}$  matter Lagrangian. It posses a global scale symmetry $g_{\mu\nu}\rightarrow e^{2w_0}g_{\mu\nu}$ in the case where the matter fields are conformally coupled to gravity ${\cal{L}}_{\rm matter}\rightarrow e^{-4w(x^{\mu})}{\cal{L}}_{\rm matter}$ and in particular in the case where $w(x^{\mu})=w_0=$constant \cite{Kounnas:2014gda}.
The equations of motions which follow from the variation of 
(\ref{S}) are:
\begin{eqnarray}
R\, R_{\mu\nu}-\frac{1}{4}R^2\, g_{\mu\nu}-\nabla_\mu\nabla_\nu R
+g_{\mu\nu}\nabla^2 R=4 \mu^2T_{\mu\nu}  \label{eq} \, ,
\end{eqnarray}
where $T_{\mu\nu}$ is the energy-momentum tensor.  It is easy to see that if we define 
\begin{eqnarray}
J_{\mu\nu}=R\, R_{\mu\nu}-\frac{1}{4}R^2\, g_{\mu\nu}-\nabla_\mu\nabla_\nu R
+g_{\mu\nu}\nabla^2 R  \label{J}\, ,
\end{eqnarray}
and utilising the identities,
\begin{eqnarray}
&&\nabla^\mu R_{\mu\nu}=\frac{1}{2} \nabla_\nu R\, , ~~~\nonumber \\
&&\nabla_\mu\nabla_\nu \nabla_\lambda R-\nabla_\nu\nabla_\mu\nabla_\lambda R=-{R^\kappa}_{\lambda\mu\nu}\nabla_\kappa R\, ,
\end{eqnarray}
we find that $J_{\mu\nu}$ is covariantly constant,
\begin{eqnarray}
\nabla^\mu J_{\mu\nu}=0\, ,
\end{eqnarray}
and therefore 
$T_{\mu\nu}$ sastisfies 
\begin{eqnarray}
\nabla^\mu T_{\mu\nu}=0\, . 
\end{eqnarray}
Our interest is to find vacuum solutions to the equations (\ref{eq}), (when $T^{\mu\nu}=0)$,
\begin{eqnarray}
R\, R_{\mu\nu}-\frac{1}{4}R^2\, g_{\mu\nu}-\nabla_\mu\nabla_\nu R
+g_{\mu\nu}\nabla^2 R=0   \label{eqv}\, .
\end{eqnarray}
It is easy to see that eq.(\ref{eqv}) is solved by two distinct classes of solutions:\footnote{The existence of these two branches is a generic feature of gravitational actions with
two or more curvature terms    \cite{Boulware:1985wk}.}

\vskip0.2cm
\noindent
(i) Flat spaces with vanishing scalar curvature:
\begin{equation}
R=0\, .
\end{equation}
As we will see in the following, this class of solutions of the $R^2$ theory contains in particular  non-Ricci flat spaces, $R_{\mu\nu}\neq 0$, which however do not exist as vacuum solutions in the dual-conformal (almost) equivalent standard Einstein gravity.

\vskip0.2cm
\noindent
(ii) Non-flat spaces which satisfy:
\begin{eqnarray}
R_{\mu\nu}=3\lambda \, g_{\mu\nu} \label{dd}\, ,
\end{eqnarray}
where $\lambda$ is an {\it arbitrary} constant\footnote{In fact 
$\lambda$ can be a function of the coordinates such that $\lambda_\mu=\partial_\mu\lambda$ is covariantly constant vector. In this case, spacetime is necessarily decomposable and eq.(\ref{dd}) leads then to $\lambda={\rm const.}$ \cite{Stephani}.}.  This is reminiscent of unimodular gravity (governed by the traceless part of Einstein equations). 
As a result, the solutions to the theory are defined up to a cosmological constant, which is nevertheless fixed by the boundary conditions. 
Of course, the limiting case with $\lambda=0$, eq.(\ref{eqv}) is also solved by Ricci-flat spaces with  $R_{\mu\nu}= 0$, as was the case of the standard Schwarzschild black hole.

\vskip0.2cm
In order to be more explicit we will study in what follows the cases of static  spherically symmetric  vacuum solutions of the form
\begin{eqnarray}
ds^2=-a(r)dt^2+b(r)dr^2+r^2 d\Omega_2^2  \label{gm}\, .
\end{eqnarray}
Then the field  equations $J_{\mu\nu}=0$ are non-linear fourth order differential equations, which are not obvious how they can be integrated in full generality. Therefore, in searching for  solution we will further assume that  $b(r)=1/a(r)$. 
Then the combination 
\begin{eqnarray}
\frac{r^2}{a}J_{tt}+J_{\theta\theta}=0 \label{eqr}
\end{eqnarray}
does not contains fourth derivatives.   Solving (\ref{eqr}) for $a^{(3)}$ and utilizing  $J_{rr}=0$ we obtain a
single equation which  is given in terms of three product  factors:
\begin{eqnarray}
 \left(r^2
   a''+4 r a'+2 a-2\right)\left(r^2 a''-2 a+2\right)\left(r a'+2 a\right) =0  \label{factor}\, .
\end{eqnarray}
The above equation has three obvious  solutions \footnote{The first two solutions of pure $R^2$ gravity already appeared in  \cite{Deser:2003up}.}:  
\begin{eqnarray}
(i)~~~~a(r)&=&1-\frac{M}{r}+\frac{K}{r^2} \label{mik} \, ,\\
(ii)~~~~a(r)&=&1-\frac{M}{r}-\lambda r^2\label{ds}\, ,\\
(iii)~~~~a(r)&=&\frac{M}{r}\label{1}\, ,
\end{eqnarray}
where $M,K,\lambda$ are independent integration constants. 
The solutions with acceptable asymptotic behavior correspond to the first and second cases. We therefore disregard the case $(iii)$.
$~$\\
$\bullet$ The vanishing of the second factor in (\ref{factor}) gives rise  to the solutions of class (ii)
discussed previously, 
\begin{eqnarray}
ds^2=-\left(1-\frac{M}{r}-\lambda r^2\right)dt^2+\frac{dr^2}{\displaystyle 1-\frac{M}{r}-\lambda r^2}+r^2 d\Omega_2^2 \label{BHds}\, .
\end{eqnarray}
When $R\neq 0$ they describe asymptotically de Sitter ($\lambda>0$) or anti-de Sitter ($\lambda<0$) spacetimes and  satisfies Eq. (\ref{dd}). For $\lambda=0$, the spacetime is asymptotically flat and they describe an ordinary black hole. 
$~$\\
$\bullet$ The vanishing of the first factor in (\ref{factor}) gives rise to the solutions of class (i) characterized by $R=0$ and $R_{\mu\nu} \neq0$,  
\begin{eqnarray}
ds^2=-\left(1-\frac{M}{r}+\frac{K}{r^2}\right)dt^2+\frac{dr^2}{\displaystyle 1-\frac{M}{r}+\frac{K}{r^2}}+r^2 d\Omega_2^2 \label{BH}\, ,
\end{eqnarray}
describing more  general  asymptotically flat solutions   which are reduced to the ordinary  Scharzschild black holes for $K=0$. In the next subsection we will investigate the scalar-flat solutions in more details.

\subsection{Non-trivial scalar-flat ($R=0$) spherically symmetric backgrounds}

A general static spherically symmetric spacetime has a metric which can  be written as 
\begin{eqnarray}
  ds^2=-\lambda(r)dt^2+\frac{dr^2}{\nu(r)}+r^2 d\Omega_2^2\, . \label{dsg}
  \end{eqnarray}  
 The trace of the vacuum  field equations
\begin{eqnarray}
 J_{\mu\nu}=0  \label{jmn}
 \end{eqnarray} 
 leads to 
 \begin{eqnarray}
 \Box R=0 \label{box}\, ,
 \end{eqnarray}
 and therefore the Ricci scalar is a harmonic function. Multiplying (\ref{box}) by $R$ and integrating over space we get that 
\begin{eqnarray}
 \int d^3x \left[\nabla_i\left(\sqrt{\lambda\nu}R\partial_i R
 \right)- \sqrt{\lambda\nu}\nabla_i R\nabla_i R\right]=0\, .
 \end{eqnarray} 
  If $\nabla_i R$ vanish sufficiently rapidly at infinity, we get from the positivity of the second term that 
\begin{eqnarray}
  \nabla_i R=0\, .
  \end{eqnarray}  
  Therefore, we get that static vacuum solutions to the $R^2$ theory are backgrounds with constant Ricci scalar $R={\rm const.}$\footnote{There is a similar argument for the $R+R^2$ theory \cite{Jacobson:1993xs}.}  In particular, 
 for $\mu^2>0$, pure $R^2$ gravity has a positive action since  
\begin{eqnarray}
 S_{\rm R^2}=\frac{1}{16\mu^2}\int d^4 x\sqrt{-g}R^2\geq 0 \, .\label{bb}
 \end{eqnarray} 
 Therefore, 
backgrounds with $R=0$ saturate the above bound  and are global minima of (\ref{bb}). For $\mu^2<0$, of course such curvature scalar-flat spacetimes are global maxima. In both cases, scalar-flat spacetimes are extrema of (\ref{bb}) and it can trivially be checked that indeed they satisfy the vacuum field equations   (\ref{jmn}). 
For a general static spherically symmetric spacetime with  metric 
of the form (\ref{dsg}), 
the curvature scalar is
\begin{eqnarray}
R= -\left(\frac{\lambda '(r)}{2 \lambda (r)}+\frac{2}{r}\right)\nu'(r)
+\left(\frac{\lambda
   '(r)^2}{2 \lambda (r)^2}-\frac{2}{r^2}-\frac{\lambda ''(r)}{\lambda (r)}-\frac{2 \lambda '(r)}{r \lambda (r)}\right)\nu(r)+\frac{2}{r^2}\, .
\end{eqnarray}
Hence, the equation  $R=0$ is written as 
\begin{eqnarray}
\nu'(r)+p(r)\nu(r)+q(r)=0\, , \label{M}
\end{eqnarray}
where 
\begin{eqnarray}
&&p(r)=\frac{2 r^2 \lambda  \lambda ''-r^2 \lambda
   '^2+4 r \lambda  \lambda '+4 \lambda ^2}{r^2
   \lambda  \lambda '+4 r \lambda ^2}, \nonumber \\
   &&~~~q(r)=-\frac{4 \lambda }{r^2 \lambda '+4 r \lambda },
\end{eqnarray}
and $\lambda(r),\nu(r)$ are subject to the conditions
\begin{eqnarray}
\lambda(r)\to 1, ~~~\nu(r)\to 1, ~~~\mbox{for $r\to \infty$}\, .
\end{eqnarray}
The general solution to (\ref{M}) is then 
\begin{eqnarray}
\nu(r)=\frac{1}{u(r)}\int u(r)q(r) dr, ~~~~u(r)=e^{\int p(r) dr}\, .
\end{eqnarray}
In other words, there is no corresponding Birkhoff theorem in pure 
$R^2$ theory: given any particular profile function $\lambda(r)$ (or $\nu(r)$), the other is uniquely determined by the asymptotic conditions and the singularity structure. For example, by choosing 
\begin{eqnarray}
\lambda(r)=\left(1-\frac{M}{r}\right)^2 
\end{eqnarray}
we find that 
\begin{eqnarray}
\nu(r)=1+\frac{\left(M-2 c_1\right) (M-2 r)}{2 r^2}\, .
\end{eqnarray}
However, it is easy to see that 
\begin{eqnarray}
C^{\mu\nu\rho\sigma}C_{\mu\nu\rho\sigma}=\frac{3 \left(-4 M^2 \left(c_1+r\right)+M r \left(8 c_1+3 r\right)-2
   c_1 r^2+2 M^3\right){}^2}{r^8 (M-r)^2} \label{c2}
\end{eqnarray}
and therefore there are singularities at $r=0$ and $r=M$. Nevertheless, the second singularity at $r=M$ is removed  for $c_1=-M/2$. For this particular value of $c_1$  we find
\begin{eqnarray}
\nu(r)=\left(1-\frac{M}{r}\right)^2 
\end{eqnarray}
so that 
\begin{eqnarray}
 ds^2=-\left(1-\frac{M}{r}\right)^2dt^2+\frac{dr^2}{\left(1-\frac{M}{r}\right)^2}+r^2\d\Omega^2_2 \label{s1}\, ,
 \end{eqnarray} 
which is just the extremal RN black hole. However, there are also other  asymptotically 
flat solutions with the correct Newtonian limit. 
For example, we find that for
\begin{eqnarray}
\lambda(r)=\left(1-\frac{M}{r}\right)^4
\end{eqnarray}
the solution of (\ref{M})
is 
\begin{eqnarray}
\nu(r)=\left(1-\frac{M}{r}\right)^{-2}
\left(1+\frac{M^2}{3 r^2}-\frac{10 M}{9 r}+\frac{c_1 e^{-\frac{3 M}{r}}}{r}\right)\, .
\end{eqnarray}
Note that there is a singularity at $r=0$ and a second one at $r=M$ as can be seen by a direct calculation of a the scalar invariant (\ref{c2}) for example. This second singularity can be removed for $c_1=-2Me^3/9$ and hence the metric is written as  
\begin{eqnarray}
 ds^2=-\left(1-\frac{M}{r}\right)^4dt^2+\frac{\left(1-\frac{M}{r}\right)^2}{1+\frac{M^2}{3r^2}-\frac{10M}{9r}\left(1+\frac{1}{5}e^{3(1-\frac{M}{r})}\right)}dr^2+r^2\d\Omega^2_2 \label{s2}
 \end{eqnarray} 
is also a vacuum static spherically symmetric solution of the $R^2$ theory. Note that the metric (\ref{s2}) has no singularity at $r=M$ as the geometry there is, like in (\ref{s1}), $AdS_2\times S^2$.
Another class of solutions is provided for example for  
\begin{eqnarray}
\lambda=e^{-M/r}\, .
\end{eqnarray}
In this case, we find that 
\begin{eqnarray}
\nu(r)= \frac{r^2 \left[r^2 \left(c_1 e^{M/r}+1024
   r\right)+4 M^3+72 M^2 r+456 M r^2\right]}{(M+4 r)^5}\, ,
\end{eqnarray}
which is asymptotically flat and singular only at $r=0$.
Therefore, the vacuum solutions of the $R^2$ theory are scalar flat $R=0$ backgrounds. In particular, these scalar flat backgrounds are global minima of the action and clearly there is no  corresponding Birkhoff theorem for the $R^2$ theory. 

\subsection{Unrevealing the no-scale  mode}

The action
\begin{eqnarray}
S=\int d^4 x\sqrt{-g}\left(\frac{1}{16 \mu^2}R^2\right)\, ,
\label{S1}
\end{eqnarray}
can equivalently be written as 
\begin{eqnarray}
S=\int d^4 x \sqrt{-g}\left(\Phi R-4\mu^2 \Phi^2\right) \, .
\label{S2}
\end{eqnarray}
The no-scale mode $\Phi$ plays the role of a Lagrange multiplier and arises in this conformal frame (Jordan) frame without space-time derivatives. In that frame the initial scale symmetry of the $R^2$ theory is translated to a  scale symmetry acting on $\Phi\rightarrow e^{-2w_0}\Phi $ compensating the scale transformation of the metric  $g_{\mu\nu}\rightarrow e^{2w_0} g_{\mu\nu}$. 
\vskip 0.2cm
Parametrizing $\Phi$ as 
\begin{equation}
\Phi=\frac{1}{2}e^{a\phi}\, , ~~~ a=\sqrt{2/3} \label{t1}
\end{equation}
and  performing a conformal transformation $g_{\mu\nu}\to \bar g_{\mu\nu}=e^{a\phi} g_{\mu\nu}$, the action (\ref{S2}) can  be written as  
\begin{eqnarray}
S=\int d^4 x \sqrt{-\bar g}\left(\frac{1}{2}\bar R-\frac{1}{2}\partial_\mu \phi \partial_\nu \phi -\mu^2\right) \label{E}\, ,
\end{eqnarray}
which shows that the initial $R^2$ action is  conformally equivalent to a conventional Einstein action coupled to an additional massless scalar propagating field $\phi$  and with a non-zero cosmological constant $\mu^2$.  
\vskip 0.2cm
Before proceeding further we would like to stress  at this point, that  the pure $R^2$ theory is scale invariant only in four dimensions. 
Indeed,  considering pure $R^2$ theory in different dimensions, 
\begin{eqnarray}
S_n=\int d^n x\sqrt{-g}\left(\frac{1}{16 \mu^2}R^2\right)\, ,
\label{S11}
\end{eqnarray}
which again can equivalently be written as
\begin{eqnarray}
S_n=\int d^n x \sqrt{-g}\left(\Phi R-4\mu^2 \Phi^2\right)\, .
\label{S22}
\end{eqnarray}
Then, by a conformal transformation, 
\begin{eqnarray}
g_{\mu\nu}\to e^{a_n\phi}g_{\mu\nu}\, ,~~ e^{a_n\phi}=2\Phi ~~~{\rm with } ~~~a_n^2={4\over(n-2)(n-1)}\, \label{E12}\, ,
\end{eqnarray} 
the theory can be written in Einstein frame as:
\begin{eqnarray}
S_n=\int d^n x \sqrt{-g}\left(\frac{1}{2}\bar R-\frac{1}{2}\partial_\mu \phi \partial_\nu \phi -\mu^2e^{\frac{4-n}{2} a_n \phi}\right) \, .
\end{eqnarray}
It is then obvious that  the $R^2$ theory is  scale invariant only for $n= 4$. 
Indeed, only in four dimensions, the potential for the $\phi$ field is flat while in higher dimensions it  acquires an exponential  potential.
The scale invariant generalization of (\ref{E1}) in $n$-dimensions is 
\begin{eqnarray}
S=\int d^nx \sqrt{-g}\left(\frac{1}{16 \mu^2}R^{n/2}\right)\, .
\label{S44}
\end{eqnarray}

\vskip 0.2cm

Returning back to the four dimensional action (\ref{E}) one derives the following field equations:
\begin{eqnarray}
&&\bar R_{\mu\nu}-\frac{1}{2}\bar g_{\mu\nu} \bar R+2\mu^2\bar g_{\mu\nu}- \partial_\mu
\phi \partial_\nu \phi+\frac{1}{2}g_{\mu\nu}(\bar \nabla\phi)^2=0 \label{E1}\, ,\\
&&\bar \nabla^2\phi=0 \label{E2}\, .
\end{eqnarray}

We will like now to investigate how the solution  
(\ref{BHds}) is written as a solution to eqs.(\ref{E1}) and (\ref{E2}). 
Let us consider eq.(\ref{eqv}) 
and now introduce a field $\phi$ defined as 
\begin{eqnarray}
e^{a\phi}=\frac{1}{8\mu^2} R \label{fr}\, ,
\end{eqnarray}
so that eq.(\ref{eqv}) is written as
\begin{eqnarray}
 R_{\mu\nu}-\frac{1}{4}e^{a\phi}\,  g_{\mu\nu}-
 e^{-a\phi}\nabla_\mu \nabla_\nu  e^{a\phi}
+ g_{\mu\nu}e^{-a\phi}\nabla^2  e^{a\phi}=0   \label{eqv2}\, .
\end{eqnarray}
Then by a conformal transformation 
\begin{eqnarray}
g_{\mu\nu}=e^{2 b\phi}\bar g_{\mu\nu}\, ,
\end{eqnarray}
and using the fact that 
\begin{eqnarray}
&&e^{-a\phi} \nabla_\mu \nabla_\nu e^{a\phi}=
a\bar \nabla_\mu\bar \nabla_\nu \phi+(a-b)a\bar \nabla_\mu \phi\bar  \nabla_\nu \phi+ab\,  \bar g_{\mu\nu}(\bar \nabla\phi)^2\, ,\\
&& R_{\mu\nu}=\bar R_{\mu\nu}-2 b \bar \nabla_\mu\bar \nabla_\nu \phi+2 b^2 \bar \nabla_\mu\phi\bar \nabla_\nu \phi -b\, \bar g_{\mu\nu}\bar \nabla^2\phi-2 b^2\, \bar g_{\mu\nu}
(\bar   \nabla\phi)^2\, ,
\end{eqnarray}
we find that Eq. (\ref{eqv2}) is written as
\begin{eqnarray}
0&=&\bar R_{\mu\nu}-(2b+a)\bar \nabla_\mu\bar \nabla_\nu \phi+(2 b^2-a^2+2ab) \bar \nabla_\mu\phi\bar \nabla_\nu \phi\, ,\nonumber \\ &&
 -(b-a)\, \bar g_{\mu\nu}\bar \nabla^2\phi-(2 b^2-ab+a^2+4ab)\,\bar  g_{\mu\nu}(\bar \nabla\phi)^2-2\mu^2\bar g_{\mu\nu}e^{(a+2b)\phi} \label{eqv4}\, .
\end{eqnarray}
We choose now 
\begin{eqnarray}
a=-2b\, ,
\end{eqnarray}
so that eqs.(\ref{fr}) and (\ref{eqv4}) turn out to be
\begin{eqnarray}
&&\bar R-6b\bar \nabla ^2 \phi-6 b^2 (\bar \nabla\phi)^2=8\mu^2 \, , \label{b1}\\ 
&& \bar R_{\mu\nu}-6b^2\bar  \nabla_\mu\phi \bar \nabla_\nu\phi -3 b \, \bar g_{\mu\nu} \bar \nabla^2 \phi-2\mu^2\bar g_{\mu\nu}=0 \label{b2}\, .
\end{eqnarray}
We see that the trace of eq.(\ref{b2}) is eq.(\ref{b1}) provided 
\begin{eqnarray}
\bar \nabla^2\phi=0\, .
\end{eqnarray}
Therefore, eqs.(\ref{b1}) and (\ref{b2}) are written (for $b^2=1/6$) as
\begin{eqnarray}
&&\bar R_{\mu\nu}-\bar \nabla_\mu\phi\bar \nabla_\nu\phi-2\mu^2\bar g_{\mu\nu}=0\, ,
\label{b3}\\ &&
\bar \nabla^2\phi=0\, ,
\end{eqnarray}
which are precisely eqs.(\ref{E1}) and (\ref{E2}). The lesson from the above exercise is the following: To any vacuum solution $ g_{\mu\nu}$ of (\ref{eqv}) with scalar curvature $ R >0$, the corresponding solution ($\bar g_{\mu\nu},\phi$) of eqs.(\ref{E1}) and (\ref{E2}) is given by
\begin{eqnarray}
&& e^{a\phi}=\frac{1}{8\mu^2} R\, , \label{tr1}\\
&& \bar g_{\mu\nu}=\frac{ R}{8\mu^2}\, g_{\mu\nu}\, .\label{tr2}
 \end{eqnarray} 
Clearly, this transformation works as long as $ R\neq 0$, otherwise the transformation is singular and the solution of the $R^2$ action (\ref{S1}) is not a solution of (\ref{E}). 

For the spherical symmetric vacuum solution 
\begin{eqnarray}\label{nonricci}
ds^2=-\left(1-\frac{M}{r}+\frac{K}{r^2}\right)dt^2+\frac{dr^2}{\displaystyle 1-\frac{M}{r}+\frac{K}{r^2}}+r^2 d\Omega_2^2 \label{BH3} \, ,
\end{eqnarray}
we find that, although it is not Ricci flat, it has vanishing Ricci scalar 
\begin{eqnarray}
 R=0\, .
\end{eqnarray}
As a result, this asymptotically  flat, spherical symmetric vacuum solution
of the pure $R^2$ theory  does not exist in the dual  Einstein gravity with cosmological constant.  This can be easily seen by varying the quadratic action of $R^2$
\begin{eqnarray}
\delta S=0~\longrightarrow~~\left(R\, \delta\sqrt{-g}+ 2 \sqrt{-g} \, \delta R\right)R=0\, ,
\end{eqnarray}
which is solved either by $R=0$ or by $ \left(\delta\sqrt{-g}+ 2 \sqrt{-g} \delta R\right)=0$. In the Einstein frame only the second class of solutions are captured. The first class with $R=0$  appears singular since it gives rise to a singular conformal transformation with $\Phi=0$. Differently stated, it corresponds to the tensionless limit (i.e. strong coupling limit) of Einstein gravity  with $M_P\rightarrow 0$. Therefore the space of general solutions of  the $R^2$ theory is  that of the dual Einstein theory but including  also its tensionless limits. 
Hence we conclude that in the tensionless limit of Einstein gravity with $M_P \rightarrow0$,
there are vacuum solutions of the type eq.(\ref{nonricci})       
with non-zero Ricci tensor but with vanishing scalar curvature.
These are however not solutions of weakly coupled gravity with finite $M_P$.

This reminds us to a similar situation in the framework of tensionless string in six dimensions where one has an equation of motion which is not possible to be obtained from an action. It also reminds us to the strong coupling limit in string theory written in the sigma-model frame, with $e^{-2\phi}R$ in the limit $g^2_s=e^{2\phi}\rightarrow \infty$. 
From eq.(\ref{S44}), the additional class of solutions with  $R=0$ exist in all dimensions  with $n>2$. Therefore the extension to the tensionless solutions does not only appear in four-dimensions but it is a common property valid in all dimensions with $n>2$. 

\vskip0.2cm
The asymptotically de Sitter solutions like  eq.(\ref{BHds})  with $\lambda>0$ having $R\ne 0$ corresponds to the second class of solutions
\begin{eqnarray}
 R_{\mu\nu}=\lambda  g_{\mu\nu}\, , ~~~~ R=12 \lambda  \, ,
\end{eqnarray}
with
\begin{eqnarray}
d s^2=-\left(1-\frac{M}{r}-\lambda r^2\right)dt^2+\frac{dr^2}{\displaystyle 1-\frac{M}{r}-\lambda r^2}+r^2 d\Omega_2^2 \label{BH4}\, ,
\end{eqnarray}
and therefore, the corresponding solution of (\ref{E}) is
\begin{eqnarray}
&&e^{a\phi}=\frac{3}{2\mu^2} \lambda \, ,\\
&&d\bar s^2=\frac{3\lambda}{2\mu^2} \left\{-\left(1-\frac{M}{r}-\lambda r^2\right)dt^2+\frac{dr^2}{\displaystyle 1-\frac{M}{r}-\lambda r^2}+r^2 d\Omega_2^2\right\} \label{BHds1}\, .
\end{eqnarray}
It is easy to see that (\ref{BHds1}) satisfies 
\begin{eqnarray}
\bar R_{\mu\nu}=2\mu^2 \bar g_{\mu\nu}\, ,
\end{eqnarray}
as expected.

\vskip 0.2cm

Note that for Ricci flat solutions with $R_{\mu\nu}=R=0$, the transformation from the $R^2$ action eq.(\ref{S1}) to the Einstein frame action eq.(\ref{E}) is still possible. In this case one has also to take the limit $\mu=0$, while keeping  $e^{a\phi}=\frac{R}{8\mu^2} $ finite.

\vskip 0.2cm

Finally, one would like to know how to find the sub-class of solutions of (\ref{S1}) once a solution of  Einstein gravity eq.(\ref{E}) is known. For this, let us suppose that we have a solution $(\bar g_{\mu\nu},\phi)$ to  eqs.(\ref{E1}) and (\ref{E2}) of the Einstein theory eq.(\ref{E}). Then, it turns out that 
\begin{eqnarray}
 R=e^{-2b \phi}\left(\bar R-6b\bar \nabla^2\phi-6b^2(\bar \nabla\phi)^2\right)\, ,
\end{eqnarray}
and the corresponding particular solution to the pure $R^2$ theory will be given by
\begin{eqnarray}
 g_{\mu\nu}=\frac{8\mu^2}{ R} \bar g_{\mu\nu}\, .
\end{eqnarray}
\section{Coupling to matter}

We would like now to couple  the $R^2$ theory to matter and find the corresponding static spherically symmetric solution. The most obvious choice for matter is an electromagnetic field, the action of which is also 
scale invariant (in fact it is conformal invariant). So we will consider below the theory 
\begin{eqnarray}\label{sa}
S=\int d^4 x\sqrt{-g}\left(\frac{1}{16 \mu^2}R^2-\frac{1}{4}F_{\mu\nu}F^{\mu\nu}\right)\, .
\label{Se} 
\end{eqnarray}
The field equations which follows from the (\ref{sa}) are 
\begin{eqnarray}
&&R\, R_{\mu\nu}-\frac{1}{4}R^2\, g_{\mu\nu}-\nabla_\mu\nabla_\nu R
+g_{\mu\nu}\nabla^2 R=4 \mu^2\left(F_{\mu\rho}{F_{\nu}}^\rho-\frac{1}{4}g_{\mu\nu} F^2\right)  \label{Eeq}\\
&&\nabla_\mu F^{\mu\nu}=0 \label{max}\, .
\end{eqnarray}
For a static spherically symmetric solution we assume that the background metric is of the form
\begin{eqnarray}
ds^2=-a(r)dr^2+\frac{dr^2}{a(r)}+r^2 d\Omega_2^2\, ,
\end{eqnarray}
so that the electromagentic field turns out to be
\begin{eqnarray}
A_t=\frac{Q}{r^2}, ~~~F_{tr}=\frac{Q}{r^2}\, .
\end{eqnarray}
The field equations (\ref{Eeq}) are written then as
\begin{eqnarray}
 \Big{(}r a'+2 a\Big{)} \Big{[}r^4 a''^2+a \left(8-8 r
   a'\right)+4 r a'(r) \left(r^2 a''+2\right)-4
   a^2+8\mu^2Q^2-4\Big{]}=0\, .
 \end{eqnarray} 
 The solution to the above equation with appropriate asymptotic behavior is
 \begin{eqnarray}
 a(r)=1-\frac{M}{r}-\frac{Z}{r^2}+\frac{Q^2\mu^2}{6Z}r^2\, ,
 \end{eqnarray}
and therefore we have 
\begin{eqnarray}
&& d\bar s^2=-\left(1-\frac{M}{r}-\frac{Z}{r^2}+\frac{Q^2\mu^2}{6Z}r^2\right)dt^2+\frac{dr^2}{1-\frac{M}{r}-\frac{Z}{r^2}+\frac{Q^2\mu^2}{6Z}r^2}+r^2 d\Omega_2^2\label{sol1}\, ,\\
&&A_t=\frac{Q}{r}\, .
\end{eqnarray}
Clearly, there are no asymptotically flat solutions (since $Q^2\mu^2/Z\neq 0$) and the only solutions that are allowed are de Sitter Reissner-Nordstrom (dS-RN) for $Z<0$ and anti-de Sitter Ressner-Nordstrom type (AdS-RN) for $Z>0$.
It can straightforwardly  be verified that, due to the conformal invariance of the Maxwell Lagrangian, the dual theory is  written as
\begin{eqnarray}
S=\int d^4 x\sqrt{-g}\left(\frac{1}{2}R-\partial_\mu\phi\partial^\mu\phi-\mu^2-\frac{1}{4}F_{\mu\nu}F^{\mu\nu}\right)\, .
\label{Sed} 
\end{eqnarray}
Then since 
\begin{eqnarray}
\bar R=\frac{-2\mu^2Q^2}{Z}\, ,
\end{eqnarray}
by using the transformations (\ref{tr1},\ref{tr2}), we find that the corresponding solution in the Einstein frame is 
\begin{eqnarray}
&& d s^2=\frac{-Q^2}{4 m^2Z}\left[-\left(1-\frac{M}{r}-\frac{Z}{r^2}+\frac{Q^2\mu^2}{6Z}r^2\right)dt^2+\frac{dr^2}{1-\frac{M}{r}-\frac{Z}{r^2}+\frac{Q^2\mu^2}{6Z}r^2}+r^2 d\Omega_2^2\right]\\
&&A_t=\frac{Q}{r}\, ,
\end{eqnarray}
and it is acceptable for only for $Z<0$ (for $\mu^2>0$). In other words, $R^2$ theory coupled to electromagnetism admits solutions of dS-RN and AdS-RN type (\ref{sol1}). However, only 
(dS-RN) are solutions of the dual Einstein gravity. Again, the dual Einstein theory can only capture part of the possible solutions of the $R^2$ theory.

\section{Entropy}

According to Wald \cite{wald1}, the black hole entropy is the Noether charge associated to diffeomorphisms under the Killing vector 
which generates the event horizon of the stationary black hole. Let us note that under an arbitrary variation of the lagrangian density we have
\begin{eqnarray}
\delta(\sqrt{-g}{ L})=\sqrt{-g}J_{\mu\nu} \delta g^{\mu\nu}+
\sqrt{-g}\nabla^\mu \theta^\mu\, .
\end{eqnarray}
The field equations are then 
\begin{eqnarray}
J_{\mu\nu}=0\, ,
\end{eqnarray} and the Noether current $\theta^\mu$ is conserved 
\begin{eqnarray}
\nabla_\mu\theta^\mu=0\, ,
\end{eqnarray}
when the field equations are satisfied. 
For diffeomorphisms generated by the vector $\xi^\mu$, we have 
\begin{eqnarray}
\delta(\sqrt{-g}{L})={\cal L}_\xi(\sqrt{-g}{L})=\sqrt{-g}\nabla_\mu (\xi^\mu L)\, ,
\end{eqnarray}
so that
\begin{eqnarray}
\nabla_\mu j^\mu=-2J_{\mu\nu}\nabla^\mu\xi^\nu,
\end{eqnarray}
where 
\begin{eqnarray}
j^\mu=\theta^\mu-\xi^\mu L \, .
\end{eqnarray}
Hence, if the field equations are satisfied, the current $j^\mu$ 
obeys
\begin{eqnarray}
\nabla_\mu j^\mu=0 \label{cons}\, .
\end{eqnarray}
The conservation equation (\ref{cons}) can be written in terms of the 1-form ${\bf j }= j_\mu dx^\mu$ as 
\begin{eqnarray}
d\ast {\bf j}=0 \, ,
\end{eqnarray}
and therefore ${\bf j}$ is the dual of an exact 3-form 
\begin{eqnarray}
{\bf j}=\ast d{\bf Q} \label{Q}\, ,
\end{eqnarray}
where ${\bf Q}=\frac{1}{2}Q_{\mu\nu}dx^\mu\wedge dx^\nu$ is a two-form. Then the Noether charge $q$ associated to a diffeomorphism generated by $\xi^\mu$ in a spatial volume $\Sigma$, which is analogous to $\int_\Sigma j_\mu d\Sigma^\mu$, is given by the boundary integral   
\begin{eqnarray}
q= \int_{\partial\Sigma }Q_{\mu\nu} \, d\Sigma^{\mu\nu}\, .
\end{eqnarray}
It turns out  that when the spacetime possess a bifurcated Killing horizon,  the Noether charge associated to the Killing vector (after appropriate rescaling to have unit surface gravity) is actually the entropy ${\cal S}$ 
\begin{eqnarray}
{\cal S}=2\pi \int_{\partial\Sigma }Q_{\mu\nu} \, d\Sigma^{\mu\nu}\, .
\end{eqnarray}
For a rigorous treatment, one may consult \cite{wald1,wald2}.
\vskip 0.2cm
The vector  $\theta^\alpha$ is constructed as follows  \cite{wald2}. 
One treats  the Lagrangian as a functional of the metric and the Riemann
tensor such that 
\begin{eqnarray}
 \delta(\sqrt{-g}L)=\sqrt{-g}\left( E_g^{\alpha\beta}\delta g_{\alpha\beta}+E_R^{\alpha\beta\gamma\delta}\delta R_{\alpha\beta\gamma\delta}+E_\phi\delta \phi\right)+\sqrt{-g}\nabla_\alpha \tilde{\theta}^\alpha \, ,
 \end{eqnarray} 
where $E_g^{\alpha\beta}$ is the variation wrt the metric, $E_R^{\alpha\beta\gamma\delta}$ is the variation with respect to to the Riemann tensor,  and  $E_\phi$ is the variation with respect to all the other matter fields, collectively denoted by $\phi$. In addition, $\tilde{\theta}^\alpha$ is different in general from  $\theta$, which is given by
\begin{eqnarray}
 \theta^\alpha=2E_R^{\alpha\beta\gamma\delta}\nabla_\delta \delta g_{\beta\gamma}-2 \nabla_\delta E_R^{\alpha\beta\gamma\delta}\delta g_{\beta\gamma}+\tilde{\theta}^\alpha \, .
 \end{eqnarray} 

Specializing to the  $R^2$ action (\ref{S1}),
we find,
\begin{eqnarray}
E_R^{\alpha\beta\gamma\delta}=\frac{1}{16\mu^2}R\left(g^{\alpha\beta}g^{\gamma\delta}-g^{\gamma\beta}g^{\alpha\delta}\right)\, ,
\end{eqnarray}
so that 
\begin{eqnarray}
\theta^\alpha=\frac{1}{8\mu^2}g^{\alpha\beta}g^{\gamma\delta}
\Big{(}R\nabla_\delta \delta g_{\beta\gamma}-R \nabla_\beta \delta g_{\gamma\delta}-\delta g_{\beta\gamma}\nabla_\delta R+\delta 
g_{\gamma\delta} \nabla_\beta R\Big{)}  \label{th}\, .
\end{eqnarray}
Then, for diffeomorphisms generated by $\xi^\alpha$ (so that $\delta g_{\alpha\beta}=\nabla_\alpha \xi_\beta-\nabla_\beta\xi_\alpha$) we find that the associated Nother current $j^\alpha$ (after using the field equations $J_{\alpha\beta}=0$), is given by
\begin{eqnarray}
j^\alpha=\frac{1}{16\mu^2}
\nabla_\beta\Big{(}R\nabla^{[\gamma}\xi^{\alpha]}\Big{)}+
\frac{1}{8\mu^2}
\nabla_\beta\Big{(}\xi^{[\gamma}\nabla^{\alpha]}R\Big{)}\, .
\end{eqnarray}
Therefore, $j^\alpha$ satisfies ($\nabla_\alpha j^\alpha=0$) 
and the 2-form $Q_{\alpha\beta}$ defined in Eq. (\ref{Q}) is given by 
\begin{eqnarray}
Q_{\alpha\beta}=\frac{1}{16\mu^2}\epsilon_{\alpha\beta\gamma\delta}
\Big{(}R\nabla^\alpha\xi^\beta+2 \xi^\alpha \nabla^\beta R\Big{)}.
\label{Q1}
\end{eqnarray}
Let us recall that $\xi^\alpha$ vanishes on a bifurcated killing horizon $B$ and for stationary backgrounds we have on $B$ that $\nabla_\alpha\xi_\beta=\kappa \epsilon_{\alpha\beta}$ where $\kappa$ is the surface gravity and $\epsilon_{\alpha\beta}$ is the  
 the binormal on $B$. Hence, by using (\ref{Q1})  we  find that the entropy is given by 
\begin{eqnarray}
{\cal S}= \frac{1}{2\mu^2}\int_B d^2 x\sqrt{h}R \label{entropy}\, .
\end{eqnarray}
This is exactly the entropy we would have found using the standard expression 
\begin{eqnarray}
{\cal S}= \frac{1}{4G}\int_B d^2 x\sqrt{h}\, ,
\end{eqnarray}
for the entropy in the Einstein frame and then using the  conformal transformation,
\begin{eqnarray}
g_{\mu\nu}\to e^{\alpha\phi}g_{\mu\nu}=\frac{16\pi G}{8\mu^2}R\, g_{\mu\nu}\, ,
\end{eqnarray}
to find the entropy in the $R^2$ theory. That the entropy is invariant under conformal transformations has been proven, at least for that class of transformations that approach identity at infinity in  \cite{Jacobson1,Jacobson2}.   
\vskip 0.2cm

It is interesting to notice that the entropy (\ref{entropy}) is zero for 
background which have $R=0$. These backgrounds cannot conformaly be mapped to Einstein frame. In addition since (for $\mu^2>0$)
backgrounds with $R=0$ saturate the  bound (\ref{bb})  and minimize the action, they resemble  the BPS condition and the associated zero entropy of the corresponding BPS states. 

\vskip 0.2cm

We also note that backgrounds with $R<0$ have   negative entropy.\footnote{The fact that higher curvature gravity may have negative entropy has been noticed also in \cite{Cvetic}.} The interpretation
of such backgrounds is not very clear. Note in particular that such backgrounds cannot be conformally mapped
to backgrounds with Lorentzian signature in the Einstein frame as long as $\mu^2>0$. The conformal mapping gives spacetime with wrong signature indicating a strong instability. The question is if we can still map backgrounds with $R<0$ to metrics  with Lorentzian signature  in the Einstein frame. This indeed can be done for  negative $\Phi<0$ in eq.(\ref{t1}), 
which can be parametrized as
\begin{eqnarray}
\Phi=-\frac{1}{2}e^{a\phi}\, , ~~~~e^{a\phi}=-\frac{1}{8\mu^2}R\, .
\end{eqnarray}
Then Einstein frame action turns out  to be
\begin{eqnarray}
S'=\int d^4 x \sqrt{-g}\left(-\frac{1}{2}R-\frac{1}{2}\partial_\mu \phi \partial_\nu \phi -\mu^2\right) \label{Eprime}\, ,
\end{eqnarray}
and the conformal transformation to the Einstein frame is only possible for $e^{a\phi}>0$, i.e. $R<0$. However, although the metric in the Einstein frame is of Lorentzian signature  now, gravity has been turned repulsive. Thus, the negative entropy in the $R^2$ theory associated to  backgrounds with $R<0$, corresponds to repulsive gravity in the Einstein frame. In other words, gravity is repulsive on backgrounds with negative entropy in the $R^2$ theory. This is a further  indication that $R^2$ theory, although it can be conformally transformed to Einstein frame, it is quite different form it. 

\section{Energy}

We note here that we may define a three-form ${\bf \Theta}=\frac{1}{3!}
\Theta_{\mu\nu\rho}dx^\mu\wedge dx^\nu\wedge dx^\rho$ as the dual of the 
1-form ${\boldsymbol{ \theta}}=\theta_\mu dx^\mu$ (${\bf \Theta}=\ast {\boldsymbol{ \theta}}$.) Let us then consider globally hyperbolic 
spacetimes which are asymptotically flat possessing Cauchy surfaces $C$ with single asymptotic region $\partial C$. 
Then, as has been proven in \cite{wald2}, if the vector field $\xi^\mu$ generates time evolution, a Hamiltonian exists if and only if there exists a 3-form ${\bf B}$ such that 
\begin{eqnarray}
\delta\int_{\partial C} \xi\cdot {\bf B}=\int_{\partial_C} \xi\cdot {\bf \Theta}\, .
\end{eqnarray}
In that case, the Hamiltonian is given by 
\begin{eqnarray}
H=\int_{\partial C}\Big{(}{\bf Q}-\xi\cdot {\bf B}\Big{)}\, ,
\end{eqnarray}
In particular, when $\xi^\mu$ is the asymptotic time-translation vector $k^\mu=(\partial/\partial t)^\mu$, the canonical energy is given by
\begin{eqnarray}
E=\int_{\partial C}\Big{(}{\bf Q}[k]-k\cdot {\bf B}\Big{)}\, .
\end{eqnarray}
For the $R^2$ theory, we find from eqs.(\ref{th}) and (\ref{Q1}) that for asymptotically flat spherically stationary solutions, we have ${\bf Q}={\bf \Theta}=0$ and therefore
\begin{eqnarray}
E=0\, ,
\end{eqnarray}
in accordance with  \cite{Boulware:1983td,Deser,Boulware:1985nn} (see also  \cite{Deser:2007vs}  for a more recent discussion on the energy definition in
$R^2$ theories).

\section{Conclusions}

We have discussed here classical solutions to the $R^2$  scale invariant gravity. As has been shown recently in \cite{Kounnas:2014gda}, this theory is equivalent to Einstein gravity with a  cosmological constant and a massless scalar field. We show how solutions of $R^2$ theory are mapped to corresponding solutions of General Relativity  in the case of non-zero scalar curvature. Furthermore, we have discussed scalar flat $R=0$ solutions of $R^2$ theory which cannot be mapped to a General Relativity setup. This class of solution corresponds to a tensionless limit respectively strong  limit in the conformally dual Einstein theory.
\vskip 0.2cm

 We worked out in details static spherically symmetric backgrounds and we showed that  in  $R^2$ theory there is no the equivalent of the Birkhoff theorem. We also discuss the coupling of this theory to scale invariant matter and in particular we study the coupling to a $U(1)$ Maxwell field. 
\vskip 0.2cm

We have also calculated the entropy and the energy for the classical solutions associated to the $R^2$ theory and we show that both the entropy and the energy vanish for scalar-flat backgrounds. This is 
a direct manifestation of the zero energy theorem \cite{Boulware:1983td},
which represent  an additional  motivation for studying this theory.
Although the spectrum of general four order theories is expected to be unbounded from below due to the presence of ghosts, the spectrum of the scale invariant  $R^2$ theory turns out to be bounded by $E_n\ge E_0=0$ at least at the classical level. This property is probably still valid at the quantum level of the theory as consequence of unitarity. This problem remains open and it will be interesting  to be proven in the future.

\vskip 2in

\noindent
{\bf \large Acknowledgment} 
\vskip .1in
We like to thank L. Alvarez-Gaume for discussions. We also thank S. Deser and B. Tekin for useful comments on the manuscript.
The research of A.K. was implemented under the Aristeia II Action of the Operational Programme Education
and Lifelong Learning and is co-funded by the European Social Fund (ESF) and National Resources.
A.K. is also partially supported by European Unions Seventh Framework Programme (FP7/2007-2013)
under REA grant agreement no. 329083. 
The work of C.K.  is also supported by the CEFIPRA/IFCPAR
4104-2 project and a PICS France/Cyprus. A.R. is supported by the Swiss National Science Foundation
(SNSF), project ``The non-Gaussian Universe" (project number: 200021140236).  
D.L. likes to thank the theory group of CERN for its hospitality. This research is also supported by the Munich
Excellence Cluster for Fundamental Physics ``Origin and the Structure of the Universe"
and by the ERC Advanced Grant ``Strings and Gravity" (Grant No. 32004). The work of
C.K. is partially supported the Gay Lussac-Humboldt Research Award 2014, at the Ludwig
Maximilians University and Max-Planck-Institute for Physics.

\newpage

\vskip.5in 
\noindent
{\bf Appendix }
\vskip.2in
\noindent
Here, for completeness,  we will provide the spherically symmetric solutions to the  scale invariant Weyl-Eddington action (\ref{W}). The field equations which follow from (\ref{W}) can be written as 

\begin{eqnarray}
W_{\mu\nu}=\beta J_{\mu\nu}-2 \alpha B_{\mu\nu}=0. \label{JB}
\end{eqnarray}
$J_{\mu\nu}$ was defined in Eq.  (\ref{J}) and  $B_{\mu\nu}$ is the Bach tensor given by
\begin{eqnarray}
B_{\mu\nu}=\left(\nabla^\rho\nabla^\sigma+\frac{1}{2}R^{\rho\sigma}\right)C_{\mu\rho\nu\sigma}.
\end{eqnarray}
Let us then consider the following form of the metric 
\begin{eqnarray}
ds^2=-a(r) dt^2+\frac{dr^2}{a(r)}+r^2 d\Omega_2^2.
\end{eqnarray}
Then, although the field equations (\ref{JB}) are complicated, we find that the 
combination $W_0^0-W_r^r$ is quite simple and turns out to be 
\begin{eqnarray}
W_0^0-W_r^r=\frac{a(r)}{6
   r^4} \Big{[}(72 \beta  r^2 a''(r)+r^3 (\alpha -12 \beta ) \left(r
   a^{(4)}(r)+4 a^{(3)}(r)\right)-144 \beta  a(r)+144 \beta \Big{]}=0.
\end{eqnarray}
The solution to the above equation is 

\begin{eqnarray}
a(r)=1-\frac{M}{r}+c_1 r^{\gamma_1}+c_2 r^{\gamma_2}+\Lambda r^2,
\end{eqnarray}
where $\gamma_{1,2}=\frac{\pm\sqrt{\alpha ^2-312 \alpha  \beta +3600 \beta ^2}+\alpha -12
   \beta }{2 \alpha -24 \beta }$ and $c_{1,2},M,\Lambda$ are integration constants. Plugging this solution to any other  components of $W_{\mu\nu}=0$ we find that $c_{1,2}=0$. Therefore the solution is finally
   \begin{eqnarray}
   a(r)=1-\frac{M}{r}+\Lambda r^2.
   \end{eqnarray}

\end{document}